\RequirePackage{fix-cm}
\documentclass[smallextended]{svjour3}       
\smartqed  
\usepackage{graphicx}
\usepackage{amssymb,amsmath}
\begin{document}

\title{Cryptanalysis and improvement of a semi-quantum private comparison protocol based on Bell states
}


\author{Li Xie       \and
        Qin Li*      \and
        Fang Yu      \and
        Xiaoping Lou \and
        Cai Zhang
}


\institute{Li Xie, Qin Li* \at
              School of Cyberspace Security, Xiangtan
University, Xiangtan 411105, China \\
\email{liqin@xtu.edu.cn}
           \and
           Fang Yu \at
              Faculty of Informatics, Masaryk University, Brno, Czech Republic\\
              Department of Computer Science, Jinan University, Guangzhou 510632, China\\
           \and
           Xiaoping Lou \at
           College of Information Science and Engineering, Hunan Normal University, Changsha 410081, China\\
           \and
           Cai Zhang \at
           College of Mathematics and Informatics, South China Agricultural University,
Guangzhou 510642, China
}

\date{Received: date / Accepted: date}

\maketitle

\begin{abstract}
Semi-quantum private comparison (SQPC) allows two participants with limited quantum ability to securely compare the equality of their secrets with the help of a semi-dishonest third party (TP). Recently, Jiang proposed a SQPC protocol based on Bell states (Quantum Inf Process 19(6): 180, 2020) and claimed it is secure. In this paper, we present two types of attack on Jiang's SQPC protocol. In the first type of attack, an outside eavesdropper will make participants accept a wrong result. In the second type of attack, a malicious participant will not only make the other participant accept a wrong result, but also learn the secret of the honest participant. Neither type of attack will be detected. In addition, we propose an improved SQPC protocol that can resist these two types of attack.
\end{abstract}

\section{Introduction}
\label{intro}
Quantum private comparison (QPC) allows two participants to compare their secrets by using the principles of quantum mechanics without revealing any information about the secrets. The first QPC protocol was put forward by Yang et al. \cite {yang2009efficient}, which can compare the equality of two participants' secrets by using EPR pairs. After that, various QPC protocols have been proposed for different situations \cite{yang2009secure,chen2010efficient,liu2011efficient,tseng2012new,wen2012quantum,sun2013quantum,chang2013multi,lin2013quantum,liu2013efficient,zhang2013cryptanalysis,zi2013quantum,lin2014quantum,li2014efficient,chen2014efficient,wang2014multi,ye2017quantum,chong2019circular}. However, the quantum capability of two participants in previous QPC protocols is assumed to be unlimited. Actually, quantum resources are difficult to obtain under current technology. Therefore, how to reduce the requirement of quantum devices for the participants becomes an important problem. In 2016, Chou et al. proposed the first semi-quantum private comparison (SQPC) protocol which reduced the quantum capability of two participants \cite{chou2016semi}. The SQPC protocol allows two ``classical'' participants to compare their secrets with the help of a semi-dishonest third party (TP) who may misbehave but cannot collude with ``classical'' participants. A ``classical'' participant here means one can only perform the following four operations: (1) preparing fresh qubits in the $ Z $-basis {$\left\{ |0\rangle , |1\rangle \right\}$, (2) measuring qubits in the $ Z $-basis, (3) reordering the qubits, and (4) sending or reflecting the qubits without disturbance.
	
However, both two participants of existing SQPC protocols all need quantum measurements. Until recently, Jiang presented a SQPC protocol based on Bell states \cite{jiang2020semi}, where quantum measurements are unnecessary. Jiang's SQPC protocol can resist some famous attack, such as the intercept-resend attack, the measure-resend attack, and the entangle-measure attack, but we show that it suffers two types of attack. In the first type of attack, an outside eavesdropper will make participants accept a wrong result without being detected. In the second type of attack, a malicious participant will not only make the other participant accept a wrong result, but also learn the secret of the honest participant. Then we improve Jiang's SQPC protocol to avoid these two types of attack.	
		
The rest of this paper is organized as follows. Section 2 briefly reviews Jiang's SQPC protocol \cite{jiang2020semi}. Two types of attack in Jiang's SQPC protocol are given in section 3. Section 4 presents the improved SQPC protocol. The security analysis of the proposed protocol is made in section 5. Finally, we make a conclusion in section 6.
\section{\textbf{Review of Jiang's SQPC protocol}}
\label{sec:1}
In Jiang's SQPC protocol \cite{jiang2020semi}, two participants Alice and Bob have their $ L $-bit secret $ X $ and secret $ Y $, respectively. The binary representations of $ X $ and $ Y $ are ($ x_{L},x_{L-1},...,x_{1} $) and ($ y_{L},y_{L-1},...,y_{1} $) in $ F_{2^{L}} $, respectively. Here, $ x_{i} $, $ y_{i} $ $ \in $ $ \{0, 1\}  $ and $ i = 1, 2, ... , L. $ They want to compare the equality of their secrets securely with the help of a semi-honest TP. A semi-honest TP means that TP may misbehave, but cannot collude with participants. In the following, Jiang's SQPC protocol is briefly reviewed.

\paragraph{\textit{\textbf{Preliminary:}}}Alice and Bob share a common key sequence $K$ of length $L$ in advance through the SQKD protocol \cite{krawec2015mediated}. Here, $ K_{i} $ is the $ i $th bit of $ K $, where $ K_{i}\in\left\{ 0, 1 \right\}$ and $ i=1,2,...,L $. Meanwhile, Alice prepares a $ L $-bit raw key $RA$. Here,  $RA_{i}$ is the $ i $th bit of $RA$, where $ RA_{i}\in\left\{ 0, 1 \right\}$ and $ i=1,2,...,L $. Bob generates a $ L $-bit raw key $RB$ similarly to Alice's operations.

\textbf{Step 1}: TP generates $2L$ Bell states randomly chosen from $ \left\{  \right. $$ |\varphi^+\rangle=\frac{1}{\sqrt{2}}(|00\rangle+|11\rangle) $, $ |\varphi^-\rangle=\frac{1}{\sqrt{2}}(|00\rangle-|11\rangle) $, $ |\psi^+\rangle=\frac{1}{\sqrt{2}}(|01\rangle+|10\rangle) $, $ |\psi^-\rangle=\frac{1}{\sqrt{2}}(|01\rangle-|10\rangle) $$ \left. \right\} $ and divides these qubits into two sequences $S_{A}$ and $S_{B}$, which denote the first and the second qubits of all Bell states. Here, $ S_{Ai} $ is the $ i $th qubit of $ S_{A} $, and $ S_{Bi} $ is the $ i $th qubit of $ S_{B} $, where $ i=1,2,...,2L $. Then TP sends $S_{A}$ and $S_{B}$ to Alice and Bob, respectively.

\textbf{Step 2}: When Alice and Bob receive each qubit, they randomly choose CTRL or SIFT. Specifically, if Alice chooses CTRL, Alice reflects the received qubit without disturbance. If Alice chooses SIFT, Alice computes $MA_{i}=K_{i}\oplus RA_{i} \oplus x_{i}$, where $ \oplus $ is the modulo 2 addition operation, generates a fresh qubit in the $Z$-basis according to $ MA_{i}$, and sends it to TP. For instance, if $ MA_{i}=0 $, Alice prepares $ |0\rangle$, else prepares $ |1\rangle$. Bob does operations similar to those Alice does according to his own choice, namely that he reflects the received qubit without disturbance or generates a fresh qubit according to $ MB_{i}$ and sends it to TP.

\textbf{Step 3}: After TP receives all the qubits sent by Alice and Bob, Alice and Bob announce their choices in step 2.

\textbf{Step 4}: TP performs one of the four operations on the received qubits based on Alice and Bob's choices as shown in Table \ref{tab:1}. In case 1, both Alice and Bob chose CTRL in step 2, TP performs the Bell measurement on the reflected qubits used for checking errors. In cases 2, 3 and 4, TP performs $Z$-basis measurements on the qubits returned by Alice or Bob who selected SIFT in step 2 to obtain the values of $MA_{i}$ and $MB_{i}$.

\textbf{Step 5}: TP evaluates the error rate in case 1 in Table \ref{tab:1}. If the error rate is higher than the predefined threshold, the protocol aborts; otherwise it continues.

\textbf{Step 6}: Alice and Bob tell TP $RA$ and $RB$, respectively. Here, $ RA= [  RA_{1},RA_{2},...,RA_{L}]  $ and $ RB= [  RB_{1},RB_{2},...,RB_{L}]  $. TP computes $R_{i} = MA_{i}\oplus MB_{i} \oplus RA_{i} \oplus RB_{i} $. If $R_{i} \neq 0$, TP will conclude that $X \neq Y$ and end the protocol due to

\begin{equation}
\begin{split}
R_{i} &= MA_{i}\oplus MB_{i} \oplus RA_{i} \oplus RB_{i} \\
&=(x_{i}\oplus RA_{i}\oplus K_{i})\oplus(y_{i}\oplus RB_{i}\oplus K_{i})\oplus RA_{i} \oplus RB_{i}\\
&=x_{i}\oplus y_{i}.
\end{split}
\end{equation}
Otherwise, TP sets $i = i + 1$ and repeats from the beginning of this step. If TP finds out that $R_{i} = 0$ in the end, TP will conclude that $X=Y$. Finally, TP announces the result.

\begin{table}\label{tab:1}
	\caption{Participants' actions on the qubit in each position in Jiang's protocol \cite{jiang2020semi}}
	\label{tab:1}       
	\begin{tabular}{lllllllllll}
		\hline\noalign{\smallskip}
		Case & Alice's action & Bob's action & TP's action \\
		\noalign{\smallskip}\hline\noalign{\smallskip}
		1 & CTRL & CTRL & To perform the Bell state measurement  \\
		& & &on both reflected quibts\\
	    & & &\\
		2 & CTRL & SIFT & To perform the $Z$-basis measurement on\\
		& & & the qubit sent by Bob\\
		& & &\\
		3 & SIFT & CTRL & To perform the $Z$-basis measurement on \\
		& & & the qubit sent by Alice \\
		& & &\\
		4 & SIFT & SIFT & To perform $Z$-basis measurements on the \\
        & & &qubits sent by both Alice and Bob\\
		\noalign{\smallskip}\hline
	\end{tabular}
\end{table}

\section{Two types of attack in Jiang's SQPC protocol }
\label{sec:4}

In this section, we will analyze two types of attack in Jiang's SQPC protocol \cite{jiang2020semi}, namely outside attack and participant attack. In the first case, an outside eavesdropper can make participants accept a wrong result. In the second case, a malicious participant can make the other participant accept a wrong result and obtain the secret of the honest participant. In both cases, the attack would not be detected. Two cases are given in the following subsections, respectively.

\subsection{\textbf{Outside attack on Jiang's SQPC protocol}}
\label{sec:5}

In this case, an outside eavesdropper Eve can intercept transmitted qubits and replace them with his own qubits in some steps to cause participants to obtain a wrong result. The specific operations of the outside attack are as follows.

In step 1, Eve intercepts $ S_{Ai} $ and $ S_{Bi} $ which TP sends to Alice and Bob and retains them in his quantum memory, then sends fake qubits to Alice and Bob, the state of which are randomly chosen from $\{|0\rangle,|1\rangle\}$.

In step 2, no matter which operation Alice and Bob chose, Eve intercepts the qubits sent by them, and sends $S_{Ai}$ and $ S_{Bi} $ in the corresponding positions retained in his quantum memory to TP.

Because the Bell state used for detection is the same as the initial Bell state prepared by TP, Eve can pass the eavesdropping check in step 5. Then, TP announces the result. But due to Eve's attack, TP obtains wrong $MA_{i}$ and $MB_{i}$ in step 4 which may lead to an incorrect result. Therefore, although Eve cannot obtain the secret message, he successfully performs an outside attack to make two participants get a wrong result without being detected.

\subsection{\textbf{Participant attack on Jiang's SQPC protocol}}
\label{sec:6}

In this subsection, we assume that Bob is a malicious participant who wants to get Alice's secret. Bob may not only intercept transmitted qubits and replace them with his own qubits in some steps to cause participants to obtain a wrong result, but also perform quantum measurements to obtain the values of $MA_i$ for learning the secret of Alice. The specific operations of the participant attack are as follows.

In step 1, for each qubit $S_{Ai}$ which TP sent to Alice, Bob intercepts it and retains it in his quantum memory, then he sends a fake qubit prepared by himself to Alice.

In step 2, no matter which operation Alice chose, Bob intercepts the qubit from Alice to TP and retains it in his quantum memory, then he sends $S_{Ai}$ retained in his quantum memory to TP instead.

In step 3, Alice and Bob announce their choices in step 2. Bob performs $Z$-basis measurements on the qubits which were sent by Alice after she selected SIFT and kept by Bob to learn the values of $MA_i$.

Because the Bell states used for detection are the same as initial Bell states prepared by TP, Bob can pass the eavesdropping check in step 5. After Alice and Bob publish $ RA_{i} $ and $ RB_{i} $, TP calculates the result by using $MA_i$ and $MB_i$ and announces it. But this result may be wrong, since Bob replaced the qubits that were sent to TP by Alice in step 2 and it may make TP obtain wrong $MA_i$ in step 4. Furthermore, when Bob obtains the values of $ RA_{i} $ published by Alice in step 6, he can learn Alice's secret by calculating

\begin{equation}
\begin{split}
&MA_{i}\oplus RA_{i} \oplus K_{i}\\
=&(x_{i}\oplus RA_{i}\oplus K_{i})\oplus RA_{i} \oplus K_{i}\\
=&x_{i}
\end{split}
\end{equation}
Therefore, Bob can make the honest participant Alice accept the wrong result and he can obtain Alice's secret without being detected.
\section{The proposed improved SQPC protocol}
\label{sec:7}
In this part, we improve the SQPC protocol proposed by Jiang \cite{jiang2020semi} and the improved one can resist the proposed two types of attack. Similar to Jiang's protocol \cite{jiang2020semi}, two participants Alice and Bob want to securely compare the equality of their secrets with the help of a semi-dishonest TP. Alice and Bob have secret $ X $ and secret $ Y $, respectively. Here, $ x_{i} $ is the $ i $th bit of $ X $ and $ y_{i} $ is the $ i $th bit of $ Y $ where $ x_{i}, y_{i}\in \{0, 1\}$ and $ i = 1, 2, ... , L$. The detailed steps are given as follows.

\paragraph{\textit{\textbf{Preliminary:}}}Alice and Bob share a common key sequence $K$ of length $L$ in advance through the SQKD protocol \cite{krawec2015mediated}. Here, $ K_{i} $ is the $ i $th bit of $ K $, where $ K_{i}\in\left\{ 0, 1 \right\}$ and $ i=1,2,...,L $. Meanwhile, Alice prepares a $ L $-bit raw key $RA$. Here,  $RA_{i}$ is the $ i $th bit of $RA$, where $ RA_{i}\in\left\{ 0, 1 \right\}$ and $ i=1,2,...,L $. Bob generates a $ L $-bit raw key $RB$ similarly to Alice's operations.

\textbf{Step 1}: TP generates $2L$ Bell states randomly chosen from $ \left\{  \right. $$ |\varphi^+\rangle=\frac{1}{\sqrt{2}}(|00\rangle+|11\rangle) $, $ |\varphi^-\rangle=\frac{1}{\sqrt{2}}(|00\rangle-|11\rangle) $, $ |\psi^+\rangle=\frac{1}{\sqrt{2}}(|01\rangle+|10\rangle) $, $ |\psi^-\rangle=\frac{1}{\sqrt{2}}(|01\rangle-|10\rangle) $$ \left. \right\} $ and divides these qubits into two sequences $S_{A}$ and $S_{B}$, which denote the first and the second qubits of all Bell states. Then TP sends $S_{A}$ and $S_{B}$ to Alice and Bob, respectively.

\textbf{Step 2}: When Alice and Bob receive each qubit, they randomly choose CTRL or SIFT. Specifically, if Alice chooses CTRL, Alice reflects the received qubit without disturbance. If Alice chooses SIFT, Alice randomly chooses one of the following two operations, namely SIFT(calculate) and SIFT(detect):

(1) SIFT(calculate): Alice performs $Z$-basis measurement on the received qubit, and records the measurement result as $MA_{i}$. Then Alice generates a fresh qubit in the $Z$-basis according to $ MA_{i} $ and sends the fresh qubit to TP. For example, if $MA_{i} = 0$, she generates $ |0\rangle$, else generates $ |1\rangle$. Meanwhile, Alice calculates $RA_{i}' = K_{i}\oplus RA_{i} \oplus x_{i} \oplus MA_{i}$, where $ i=1,2,...,L $.

(2) SIFT(detect): Alice discards the qubit from TP and generates a trap qubit randomly chosen from $ \left\{ |0\rangle , |1\rangle \right\} $ and sends it to TP.

Bob does operations similar to that Alice does according to his own choice. He will also randomly choose CTRL or SIFT. And if he selects SIFT, there are two choices, namely SIFT(calculate) and SIFT(detect). For SIFT(calculate), he generates a fresh qubit according to the value of $ MB_{i}$ to be sent to TP and evaluates $RB_{i}' = K_{i}\oplus RB_{i} \oplus y_{i} \oplus MB_{i}$; and for SIFT(detect), he generates a trap qubit and sends it to TP.

\textbf{Step 3}: After TP receives all the qubits sent by Alice and Bob, Alice and Bob announce their choices in step 2. In addition, Alice and Bob also announce which SIFT operations are SIFT(detect) since they will be used to detect eavesdropping.

\textbf{Step 4}: TP performs different operations on the received qubits based on Alice and Bob's choices. Since both Alice and Bob can choose CTRL, SIFT(calculate), and SIFT(detect), there are nine cases as shown in Table \ref{tab:2}. In case 1, both Alice and Bob chose CTRL in step 2, TP performs the Bell measurement on the reflected qubits. In the other eight cases, TP performs $Z$-basis measurements on the qubits which were sent by Alice or Bob if they selected SIFT in step 2 to learn the values of $MA_{i}$, $MB_{i}$, and the measurement results of trap qubits. Then TP announces the measurement results of the trap qubits and Alice and Bob compare whether they are consistent with the states of the trap qubits originally prepared by themselves.

\begin{table}
	\caption{Participants' actions on the qubit in each position in the improved SQPC protocol}
	\label{tab:2}       
	\begin{tabular}{llll}
		\hline\noalign{\smallskip}
		Case & Alice's action & Bob's action & TP's action \\
		\noalign{\smallskip}\hline\noalign{\smallskip}
		1 & CTRL & CTRL & To perform the Bell state measurement \\
         & & &on both reflected quibts\\
		& & &\\
		2 & CTRL & SIFT(calculate) & To perform the $Z$-basis measurement on\\
		& & & the qubit sent by Bob\\
		& & &\\
		3 & CTRL & SIFT(detect) & To perform the $Z$-basis measurement on\\
		& & & the qubit sent by Bob\\
		& & &\\
		4 & SIFT(calculate) & CTRL & To perform the $Z$-basis measurement on\\
		& & & the qubit sent by Alice\\
		& & &\\
		5 & SIFT(detect) & CTRL & To perform the $Z$-basis measurement on\\
		& & & the qubit sent by Alice\\
		& & &\\
		6 & SIFT(calculate) & SIFT(detect) & To perform $Z$-basis measurements on the\\
        & & &qubits sent by both Alice and Bob\\
		& & &\\
		7 & SIFT(detect) & SIFT(calculate) & To perform $Z$-basis measurements on the\\
& & &qubits sent by both Alice and Bob\\
		& & &\\
		8 & SIFT(calculate) & SIFT(calculate) & To perform $Z$-basis measurements on the\\
& & &qubits sent by both Alice and Bob\\
		& & &\\
		9 & SIFT(detect) & SIFT(detect) & To perform $Z$-basis measurements on the\\
& & &qubits sent by both Alice and Bob\\
		\noalign{\smallskip}\hline
	\end{tabular}
\end{table}

\textbf{Step 5}: TP evaluates the error rate in case 1 in terms of the results of the Bell measurements and Alice and Bob evaluates the error rate in cases 3, 5, 6, 7, and 9 according to the measurement results of trap qubits. If any one error rate is higher than the predefined threshold, the protocol aborts; otherwise it continues.

\textbf{Step 6}: Alice and Bob publish the value of $RA_{i}\oplus RA_{i}'$ and $RB_{i}\oplus RB_{i}'$ where $ i=1,2,...,L $, respectively. TP computes
\begin{equation}
\begin{split}
R_{i} &= MA_{i}\oplus MB_{i} \oplus (RA_{i} \oplus RA_{i}') \oplus (RB_{i} \oplus RB_{i}')\\
      &= MA_{i}\oplus MB_{i} \oplus (RA_{i} \oplus K_{i} \oplus RA_{i} \oplus x_{i} \oplus MA_{i})\\
      &\oplus(RB_{i} \oplus K_{i} \oplus RB_{i} \oplus y_{i}\oplus MB_{i})\\
      &=  x_{i}\oplus y_{i}.
\end{split}
\end{equation}
If $R_{i} \neq 0$, TP will conclude that $X \neq Y$ and end the protocol. Otherwise, TP sets $i = i + 1$ and repeats from the beginning of this step. If TP finds out that $R_{i} = 0$ in the end, TP will conclude that $X=Y$. Finally, TP announces the result.

\section{Security analysis}
\label{sec:10}

In this section, we show the improved SQPC protocol can avoid the proposed two types of attack on the SQPC protocol proposed by Jiang \cite{jiang2020semi} and also can resist other types of attack as Jiang's protocol did.

Firstly, we show that the proposed outside attack on the SQPC protocol in Ref. \cite{jiang2020semi} is impossible in the proposed protocol. In step 1, the outside eavesdropper Eve intercepts the qubits sent by TP and retains them in his quantum memory. In step 2, Eve intercepts the qubits returned by two participants. In these two steps, the participants will be unaware of Eve's eavesdropping. However, in step 3, as the participants announced the positions of the trap qubits to be detected, TP should measure these trap qubits and publish the measurement results in step 4. If Eve forged these qubits and returned to TP, the probability that makes the measurement result of each trap qubit in an error is $ \frac{1}{2} $ if either Alice or Bob chose SIFT(detect). Thus the probability that Eve is detected will be $ 1-(\frac{1}{2})^{n+m} $, where $ n $ and $ m $ are the number of the trap qubits prepared by Alice and Bob, and it is reduced to 1 when $ n $ and $ m $ are sufficiently large.

Secondly, we show that the proposed participant attack on the SQPC protocol in Ref. \cite{jiang2020semi} is also invalid in the proposed protocol. Suppose Bob is dishonest. In steps 1 and 2, Bob performs similar operations as the previous outside eavesdropper did. The participants should announce the positions of the trap qubits in step 3 and TP will measure these trap qubits and publish the measurement results in step 4. If Bob forged the qubits returned by Alice, the probability that TP produces a wrong measurement result of the trap qubit is $ \frac{1}{2} $ and the probability of Bob passing the security check is also $ \frac{1}{2} $ if Alice chooses SIFT(detect). Therefore, the probability of the dishonest participant being detected will be $ 1-(\frac{1}{2})^{n} $, where $ n $ is the number of the trap qubits produced by Alice, and it is close to 1 if $ n $ is large enough.

Except the proposed two types of attack, the improved SQPC protocol also can resist some other typical types of attack such as the intercept-resend attack, the measure-resend attack, the entangle-measure attack, and semi-honest TP attempting to learn secret information, as the original Jiang's protocol did. The reason is that the basic steps of the two protocols are similar and the main difference mainly lies in that in step 2 of the proposed protocol, the SIFT operation is divided into two cases. In one case, SIFT(detect) is to detect attack by trap qubits and in the other case SIFT(calculate) requires the participants to measure the received qubits instead of discarding them directly like Jiang's protocol \cite{jiang2020semi}. Thus we only briefly analyze these types of attack in the following and more details can be found in Ref. \cite{jiang2020semi}.

For the intercept-resend attack, an eavesdropper Eve may intercept the qubits during the transmission and send fake qubits to two participants or TP. In step 2 of the proposed protocol, if both Alice and Bob chose CTRL, the initial Bell state is not changed because they reflected the received qubits without disturbance. However, Eve does not know the choices of Alice and Bob. If he sends fake qubits to the participants in this case, TP can detect Eve's attack by performing the Bell measurements on them in step 4.

The measure-resend attack here is that Eve intercepts the qubits and measures them in the $Z$-basis to obtain some information and then sends the qubits to two participants or TP. In step 4, TP will perform the Bell measurement on the received qubits which were sent by Alice and Bob if they selected CTRL in step 2. Eve will also be detected by TP because the measurements made by Eve will destroy the initial Bell states.

The entangle-measure attack means that Eve entangles each transmitted qubit with an ancillary qubit and then performs measurements on the ancillary qubits later with the intention of obtaining some information. According to Ref. \cite{jiang2020semi}, if Eve does not want to induce any errors in step 5, no matter what Alice or Bob chose in step 2, his ancillary qubits should not entangle with two participants' qubits. Thus Eve cannot obtain any useful information by performing measurements on the ancillary qubits.

In addition, although the semi-honest TP must faithfully execute the procedure of the protocol, he may attempt to deduce the secret information of the two participants based on the public information and the records of all intermediate calculations. TP can get $MA$ and $MB$ by performing measurements and $RA \oplus RA'$ and $RB\oplus RB'$ which are published by the two participants, but he still cannot obtain the information about $X$ and $Y$ because the key $K$ shared by Alice and Bob is unknown to TP.

\section{Conclusion}
\label{sec:14}
In conclusion, we have proposed two types of attack based on Jiang's SQPC protocol \cite{jiang2020semi}. In the first type of attack, an outside eavesdropper can make two participants accept a wrong result. In the second type of attack, a dishonest participant can make the other participant accept a wrong result and obtain the secret of the other participant. Neither type of attack will be detected. Furthermore, we have proposed an improved SQPC protocol which can avoid these two types of attack just by increasing the quantum measurement capability of the participants. Actually, another solution to resist the proposed two kinds of attack is to enable ``classical'' participants to have quantum memory to reorder qubits, but quantum memory is more difficult to realize in current technologies. 

\begin{acknowledgements}
This work was supported by the Joint Funds of the National Natural Science Foundation of China and China General Technology Research Institute (Grant No. U1736113), the Key Project of Hunan Province Education Department (Grant No. 20A471), the Natural Science Foundation of Hunan Province (Grant No. 2018JJ2403), and the National Natural Science Foundation of China (Grant No. 61902132).
\end{acknowledgements}

%
%




\end{document}